\documentclass{article}
\usepackage[left=2.5cm,right=2.5cm,top=2.5cm,bottom=2.5cm,bindingoffset=0cm]{geometry}
\usepackage{cmap}
\usepackage[english]{babel} 
\usepackage{indentfirst}
\usepackage{amssymb,amsmath}
\usepackage{graphicx}
\usepackage[hyphens]{url} 
\usepackage{amsthm}
\usepackage{algpseudocode} 
\usepackage{pgfplots}
\usepackage{float}
\usepackage{hyperref}

\theoremstyle{definition}

\title{Series Resonant Matrix Converter Topology for EV DC Fast Charging}
\date{}
\author{Alex Borisevich, PhD, \href{mailto:akpc806b@gmail.com}{akpc806b@gmail.com}}

\DeclareMathOperator{\const}{const}

\begin{document}

\maketitle

\section*{Abstract}

In this paper, we presented a matrix converter with a high-frequency DC link for EV DC fast charging applications. Introducing this topology allows eliminating electrolytic capacitors in DC link of the converter, which potentially increases the reliability and power density of the system. The converter uses bidirectionally blocking three-phase rectifier to generate a high-frequency square wave voltage for the series resonant LC DC/DC converter, eliminating one H-bridge in front of the resonant tank. Principles of high-frequency and low-frequency modulations are described. A low voltage prototype has been built and tested which demonstrated ZVS soft switching of all the transistors in the converter. Peak efficiency estimated as 98.0 \% for state-of-the-art switches.

\section{Converter Topology}

\begin{center}
\hspace*{-1cm}
\ifpdf 
  \includegraphics[width=1.15\textwidth]{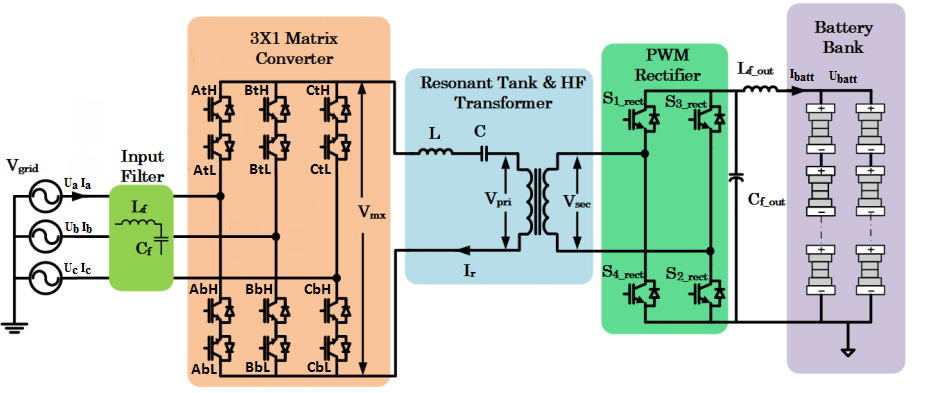}
\fi

Figure 0. Proposed series resonant matrix converter topology for EV battery charging.

\end{center}

The proposed topology for EV battery charging uses a high frequency transformer for power processing as shown in Figure 0.
The topology can be divided into three stages: 3x1 matrix converter (MC), LC series resonant tank (SRC) and high frequency (HF) transformer, and a full bridge synchronous rectifier. 

The three phase utility grid is interfaced to the MC via an Lf-Cf filter. The MC consists of six bidirectional MOSFET switches (S11-S23) as shown in Figure 0. The matrix converter stage takes a line frequency three phase sinusoidal input voltage and delivers a high frequency square wave output. 

An LC resonant tank is connected across the output of the MC in series with the HF transformer primary. The output power transferred through the transformer is regulated by changing a frequency of square wave voltage applied to the resonant tank. The operating frequency range lies above resonant frequency of LC network in order to achieve ZVS turn ON conditions.

The LC resonant tank is interfaced to a full bridge synchronous rectifier via the HF transformer. The rectifier is performing synchronous rectification of the AC voltage in order to produce DC current applied to the battery. Additionally to the switching frequency regulation, the DC output current can be further constrained by introducing either phase shift or variation of duty cycle for rectifier bridge.

Topologies of this kind (matrix converter followed by series resonant tank) were described in [1,2]. Additionally, matrix converters separately have been reviewed in [3]. Worth noting that 3x1 matrix converters are considered as a part of indirect matrix converters [4], and proposed topologies are also considerably popular for wireless power transfer [5].

\section{Principle of operation}

The battery at every moment of time $t$ is being charged with commanded from BMS current $I_{batt}$ and having voltage $V_{batt}$ at its terminals. We can assume that $I_{batt}$ and $V_{batt}$ are slowly moving and are quasi-constant during one AC grid period. Thus the converter should deliver a constant power (in a time scale of single AC grid period):

\begin{equation}
P_{batt} = V_{batt} I_{batt} = \const
\end{equation}

\subsection{Matrix converter modulation strategy}

The commutation process of the matrix converter is following these three goals:

1. Selecting appropriate input phases to maximize the amplitude of square wave applied to the resonant tank and also achieve unity input power factor. Let's name it low frequency PWM pattern. In order to generate low frequency PWM pattern, a part of the matrix converter is considered as a voltage DC-Link PWM inverter (following from [3]).

2. Inside the low frequency PWM pattern, an additional high frequency PWM pattern should be inserted in order to generate a high frequency square wave with alternating polarity to feed the resonant tank. A pair of phases which will supply energy to the resonant tank is defined by the low frequency PWM pattern, thus the high frequency pattern just recreates a PWM modulation used to drive input H-bridge of LLC or LC converter. In contrast with low frequency PWM pattern, the frequency of this switching should be variating in order to regulate the power transferring through the resonant tank.

3. High frequency PWM pattern also should consist of dead time where freewheeling resonant tank current will recharge drain-to-source capacitance of switched devices, and after that, it will flow through the body diodes ensuring ZVS turn ON switching.

\subsection{Low frequency PWM modulation}

The low frequency PWM modulation is needed to achieve unity input power factor, i.e. to have phase currents proportional to phase voltages. 

The whole AC grid period can be divided into 6 sectors depending on a polarity of input voltages. The input sectors are defined as shown in Figure 1. The six input sectors are shown in Figure 1 can be classified into two cases: 

- In the first case, one input voltage is positive and two input voltages are negative (sectors 1, 3, 5). It is obvious, that the phase with positive input voltage has maximum absolute voltage over the other two phases, in order to maximize applied voltage amplitude to the resonant tank. Thus this phase can be used as a positive terminal of virtual DC source supplying resonant tank. The negative source terminal is being selected among two remaining phases with accordance to low frequency switching pattern.

- In the second case, two input voltages are positive and one input voltage is negative (sectors 2, 4, 6). It is obvious, that the phase with negative input voltage has maximum absolute voltage over the other two phases. Thus this phase can be used as a negative terminal of virtual DC source supplying resonant tank. The positive source terminal is being selected among two remaining phases with accordance to low frequency switching pattern.

\begin{center}
\ifpdf 
  \includegraphics[width=1.0\textwidth]{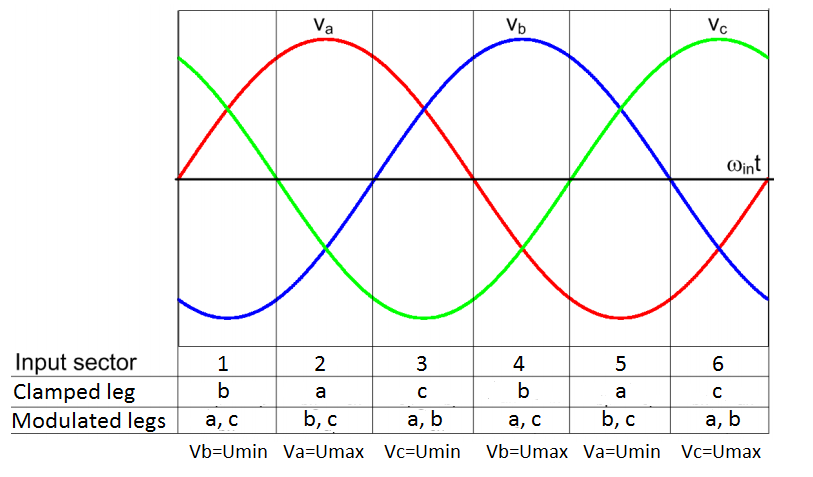}
\fi

Figure 1. The definition of input sectors.

\end{center}

To explain the principle of low frequency PWM pattern generation, we can consider without loss of generality the rectifier operation in sector 2. In this sector, the $V_a$ is being used all the time as an instantaneous positive supply voltage terminal, while $V_b$ and $V_c$ are negative and commutated to the load by PWM modulation.

Following [4], the duty cycles for switches in phases $a$, $b$ and $c$ are being calculated as:

\begin{equation}\label{eq:low_freq_pwm}
d_a = 1, \; d_b = -\frac{V_b}{V_a}, \; d_c = -\frac{V_c}{V_a}
\end{equation}

Under the balanced input voltages, the average DC-link voltage amplitude at the output of matrix converter is obtained as follows [4]:

\begin{equation}\label{eq:v_dc}
V_{dc}(t) = \frac{3}{2} \cdot \frac{V_{in}^2}{V_a(t)}
\end{equation}

where $V_{in}$ is amplitude of input phase to neutral voltages.

Let's suppose battery is demanding constant power $P_{batt}$ during the AC grid period, and high frequency switching for resonant tank is being regulated in order to maintain constant power $P_{batt}$ output with variable instantaneous input voltage $V_{dc}(t)$.

Thus the average DC-link current is being calculated as:
\begin{equation}\label{eq:i_dc}
I_{dc}(t) = \frac{P_{batt}}{V_{dc}(t)} = \frac{2 P_{batt}}{3 V_{in}^2} \cdot V_a(t) = k \cdot V_a(t)
\end{equation}

where $k = 2 P_{batt} / (3 V_{in}^2)$ is a proportional constant.

Note that the average phase input currents are proportional with duty cycles to DC-link current $I_{dc}$ (with a sign corresponding to commutated polarity):

\begin{equation}
I_a = d_a I_{dc}, \; I_b = - d_b I_{dc}, \; I_c = -d_b I_{dc}
\end{equation}

Then taking into account \eqref{eq:i_dc} and \eqref{eq:low_freq_pwm}, we can formally deduce sinusoidal form of input current, i.e. unity power factor:

\begin{equation}
\begin{gathered}
I_a = d_a I_{dc} = I_{dc} = k \cdot V_a \\
I_b = -d_b I_{dc} = \frac{V_b}{V_a} I_{dc} = k \cdot V_b \\
I_c = -d_b I_{dc} = \frac{V_c}{V_a} I_{dc} = k \cdot V_a
\end{gathered}
\end{equation}

If duty cycles $d_a, d_b, d_c$ for all three phases are calculated, then the binary switching signals $\text{pwm}_a, \text{pwm}_b, \text{pwm}_c$ will be produced by three PWM timers with fixed frequency and variable duty cycle.

\subsection{High frequency PWM modulation}

In order to define high frequency PWM pattern, let's recall driving method of well known LC or LLC converter (Figure 2). 

\begin{center}
\ifpdf 
  \includegraphics[width=0.6\textwidth]{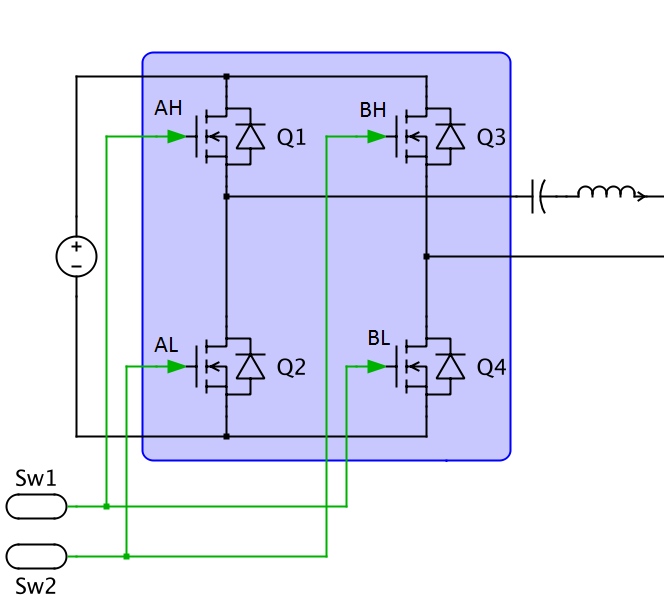}
\fi

Figure 2. Input bridge of LC converter.

\end{center}

The driving signals of two bridge legs are: $AH, AL, BH, BL$. For a particular case of fully-driven primary side of converter: $AH = BL$ and $AL = BH$ as it seen from the Figure 3. The decreasing duty cycle can be achieved by simply shifting leg B respect to the leg A. The ZVS commutation is happening during dead-time between $xH$ and $xL$ pulses (x is either A or B).

\begin{center}
\ifpdf 
  \includegraphics[width=0.9\textwidth]{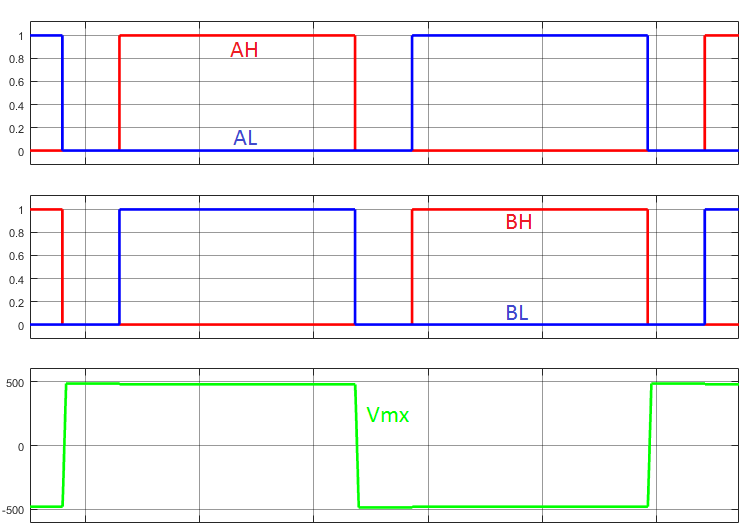}
\fi

Figure 3. PWM signals and output voltage of LC converter input bridge.

\end{center}

In order to define high frequency PWM pattern for matrix converter, we need to recognize in the matrix converter bridge (Figure 0) all the switches which are consisting the input bridge of LC converter (Figure 2) for each input sector (Figure 1) and each low frequency PWM combination.

For example lets consider input sector 2 with clamped phase a and modulated phases b and c. For this sector we have two states of low frequency PWM signals: $\text{pwm}_a = 1, \; \text{pwm}_b = 1$ and $\text{pwm}_a = 1, \; \text{pwm}_c = 1$. In the first case a-b phase-to-phase voltage $V_{ab}$ is selected to be the output of the bridge $V_{mx}$, while in the second case voltage $V_{ac}$ is being commutated to the matrix output. The corresponding circuits are demonstrated in Figure 4. 

Then we can deduce what switches are being commutated with high frequency pattern by observing in Figure 2 that the positive resonant current is always flowing from drain to source of commutated MOSFET transistors.

\begin{center}
\ifpdf 
  \includegraphics[width=1.1\textwidth]{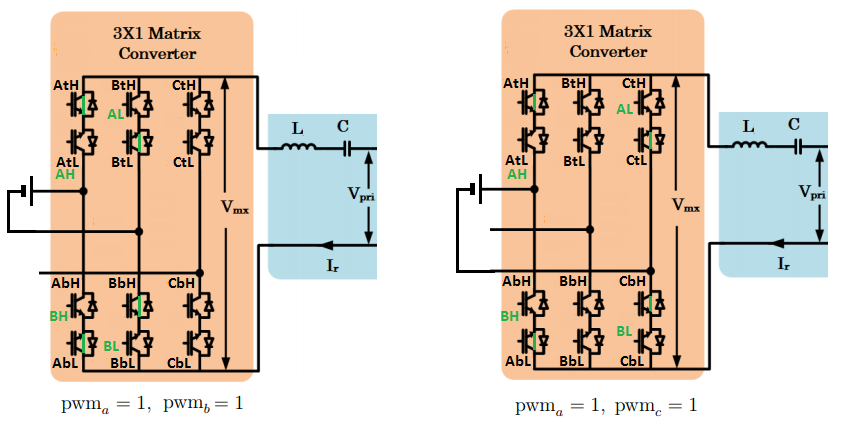}
\fi

Figure 4. Two possible cases of input voltage connection for input sector 2.

\end{center}

As we can see from the Figure 4, following commutation strategy is possible:

- For $\text{pwm}_a = 1, \; \text{pwm}_b = 1$: $AtH = 1$, $BtL = 1$, $AbL = 1$, $BbH = 1$, and $AtL = A_H$, $BtH = A_L$, $AbH = B_H$, $BbL = B_L$.

- For $\text{pwm}_a = 1, \; \text{pwm}_c = 1$: $AtH = 1$, $CtL = 1$, $AbL = 1$, $CbH = 1$, and $AtL = A_H$, $CtH = A_L$, $AbH = B_H$, $CbL = B_L$.

where $AH, AL, BH, BL$ are signals for the H-bridge switching according to the Figure 2.

Considering by analogy all the other input sectors, the following table can be produced for selecting low-frequency and high-frequency commutated switches:

\begin{table}[H]
\hspace*{-1cm}
\begin{tabular}{|l|lll|llllllllllll|}
\hline
Sector & $\text{pwm}_a$ & $\text{pwm}_b$ & $\text{pwm}_c$ & AtH & AtL & AbH & AbL & BtH & BtL & BbH & BbL & CtH & CtL & CbH & CbL \\ \hline 
1 & 0 & 1 & 1 & 0 & 0 & 0 & 0 & AL & 1 & 1 & BL & 1 & AH & BH & 1 \\
1 & 1 & 1 & 0 & 1 & AH & BH & 1 & AL & 1 & 1 & BL & 0 & 0 & 0 & 0 \\
2 & 1 & 1 & 0 & 1 & AH & BH & 1 & AL & 1 & 1 & BL & 0 & 0 & 0 & 0 \\
2 & 1 & 0 & 1 & 1 & AH & BH & 1 & 0 & 0 & 0 & 0 & AL & 1 & 1 & BL \\
3 & 1 & 0 & 1 & 1 & AH & BH & 1 & 0 & 0 & 0 & 0 & AL & 1 & 1 & BL \\
3 & 0 & 1 & 1 & 0 & 0 & 0 & 0 & 1 & AH & BH & 1 & AL & 1 & 1 & BL \\
4 & 0 & 1 & 1 & 0 & 0 & 0 & 0 & 1 & AH & BH & 1 & AL & 1 & 1 & BL \\
4 & 1 & 1 & 0 & AL & 1 & 1 & BL & 1 & AH & BH & 1 & 0 & 0 & 0 & 0 \\
5 & 1 & 1 & 0 & AL & 1 & 1 & BL & 1 & AH & BH & 1 & 0 & 0 & 0 & 0 \\
5 & 1 & 0 & 1 & AL & 1 & 1 & BL & 0 & 0 & 0 & 0 & 1 & AH & BH & 1 \\
6 & 1 & 0 & 1 & AL & 1 & 1 & BL & 0 & 0 & 0 & 0 & 1 & AH & BH & 1 \\
6 & 0 & 1 & 1 & 0 & 0 & 0 & 0 & AL & 1 & 1 & BL & 1 & AH & BH & 1 \\ \hline
\end{tabular}
\end{table}

\section{Implementation results}

\subsection{Simulation}

Matlab simulation was conducted in order to verify ZVS commutation of the switched transistors. The particular waveform of the output voltage of the matrix bridge is shown in Figure 5.

\begin{center}
\ifpdf 
  \includegraphics[width=1.0\textwidth]{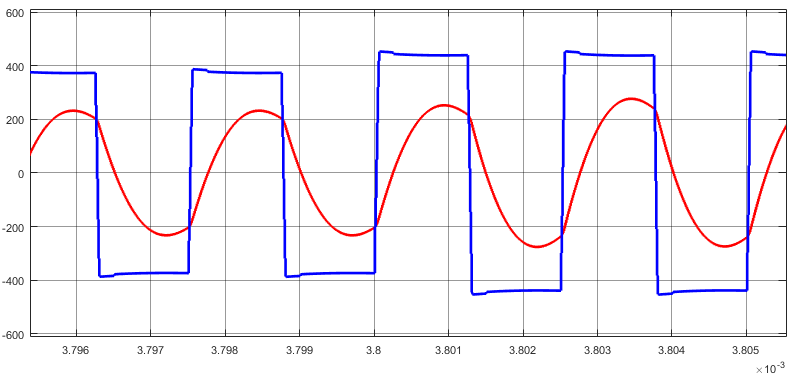}
\fi

Figure 5. Simulated output voltage (blue) and resonant tank current (red) at the output of matrix converter.

\end{center}

The low frequency modulation strategy was verified by connecting constant power load to the output of matrix converter and calculating PWM cycle average phase currents. The result is demonstrated at Figure 6.

\begin{center}
\ifpdf 
  \includegraphics[width=1.0\textwidth]{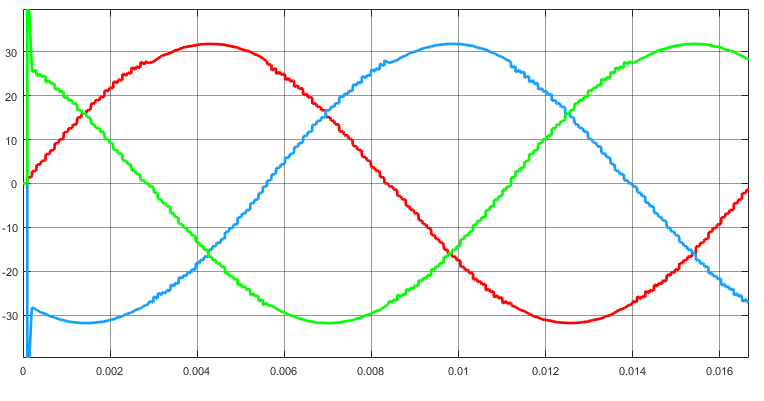}
\fi

Figure 6. Simulated input phase currents.

\end{center}

\subsection{Hardware prototype}

In order to verify feasibility of converter control implementation and prove key assumption, a hardware prototype has been made (Figure 7) for operation at low voltage. The matrix converter was loaded to RLC resonant tank which emulates a transformer followed by synchronous rectifier.

\begin{center}
\ifpdf 
  \includegraphics[width=1.0\textwidth]{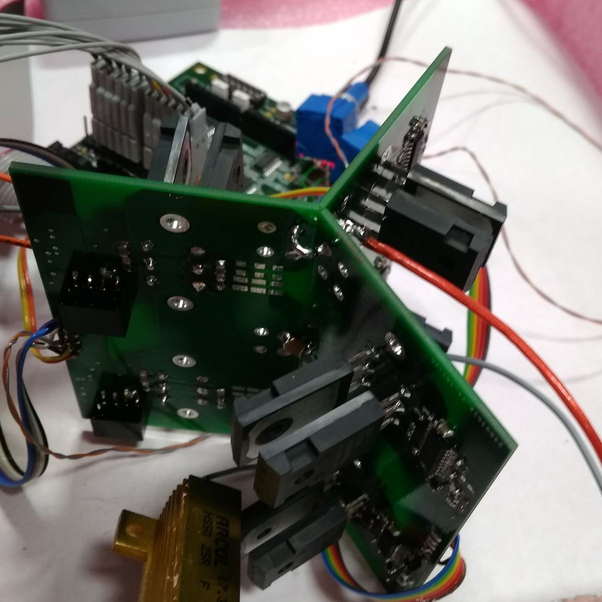}
\fi

Figure 6. Hardware prototype of resonant matrix converter.

\end{center}

The first test was conducted to ensure ZVS switching for only one pair of phase legs (the left side of Figure 4). The waveform of matrix bridge output voltage and resonant tank current presented in Figure 7 along with corresponding PWM signals. The ZVS commutation can be appreciated from a clearly visible diode conduction time (highlighted by yellow circles).

\begin{center}
\ifpdf 
  \includegraphics[width=1.0\textwidth]{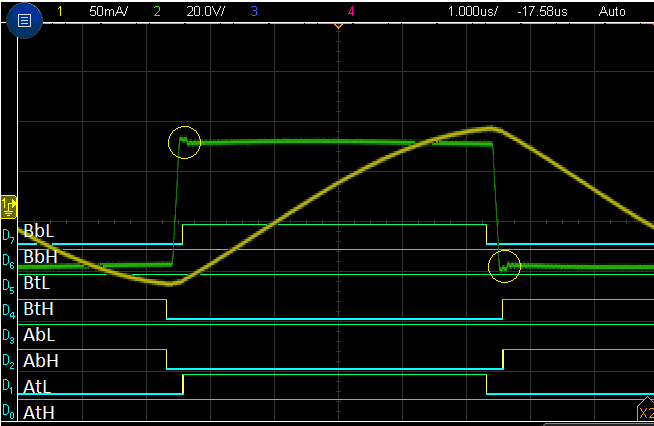}
\fi

Figure 7. Matrix converter output and resonant tank current for the sector 2 and $\text{pwm}_a = 1, \; \text{pwm}_b = 1$.

\end{center}

The second test was conducted in order to verify ZVS commutation during the switching with accordance to low-frequency PWM pattern. In this case alternation between two phase legs happens, as it pictured in Figure 4. The waveform of matrix bridge output voltage and resonant tank current presented in Figure 8 along with corresponding PWM signals. The ZVS commutation can be appreciated from a clearly visible diode conduction time (highlighted by yellow circle).

\begin{center}
\ifpdf 
  \includegraphics[width=1.0\textwidth]{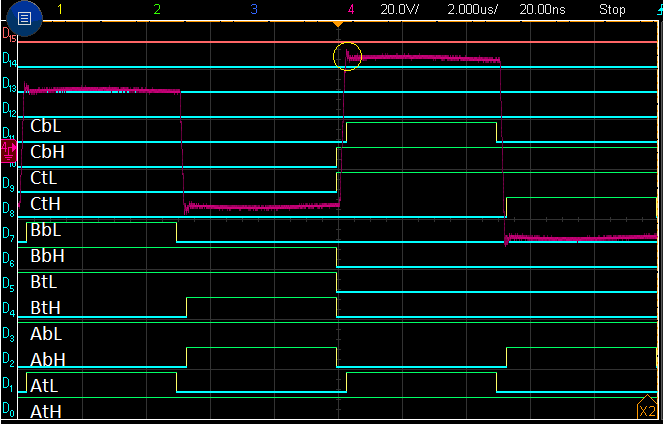}
\fi

Figure 8. Matrix converter output and resonant tank current for low-frequency transition inside the sector 2 between $\text{pwm}_a = 1, \; \text{pwm}_b = 1$ and $\text{pwm}_a = 1, \; \text{pwm}_c = 1$.

\end{center}

Afterwords, the complete implementation of PWM strategy was tested in all 6 sectors (intervals) of three-phase AC input voltages. The input voltages $V_a$ and $V_b$ along with decoded sector code and PWM signals for all transistors are demonstrated in Figure 9. 

\begin{center}
\ifpdf 
  \includegraphics[width=1.0\textwidth]{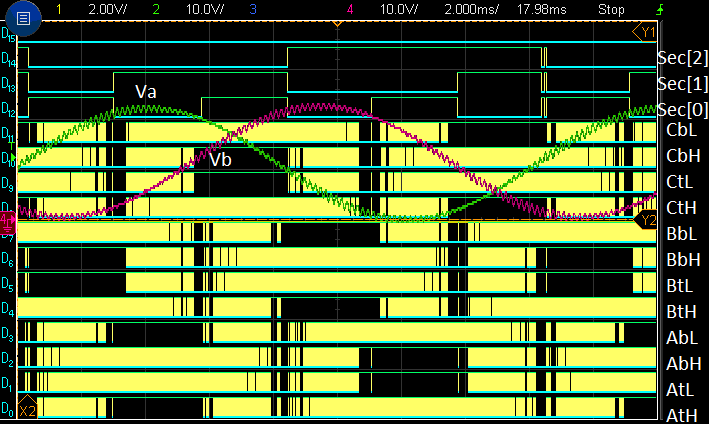}
\fi

Figure 9. PWM signals for a whole period of AC input along with decoded sector number.

\end{center}

\subsection{Efficiency estimation}

The best efficiency is being achieved at buck-to-boost operating point when $V_{pri} = n \cdot V_{sec}$, where $n$ is transformer turns ratio, and $V_{pri}$ and $V_{sec}$ are amplitude voltages at primary and secondary windings of the transformer correspondingly.

The output voltage of the matrix bridge is normally variating between $V_{pri}^{min} = 415.7$V to $V_{pri}^{max} = 480$V, so the primary voltage for turns ratio calculation can be assumed as $V_{pri} = (V_{pri}^{min} + V_{pri}^{max}) / 2$, which is $V_{pri} = 447.75$V.

The battery voltage can be assumed in boundaries of $V_{sec}^{min} = 325$V to $V_{sec}^{max} = 425$V, so the average value can be assumed as $V_{sec} = (V_{sec}^{min} + V_{sec}^{max}) / 2$, which is $V_{sec} = 375$V.

Thus, the turns ratio $n = V_{pri} / V_{sec}$ which is $n = 1.2$ for the voltages calculated above.

We can assume that if the battery voltage $V_{batt}$ is around $V_{sec}$, then there is no voltage drop across the resonant tank and there is no circulating current through the input matrix bridge between input filter and resonant tank. (This assumption is not exactly true since the output voltage of matrix converter can't be exactly $V_{pri} = n \cdot V_{sec}$ all the time and pulsating according to \eqref{eq:v_dc}). 

Considering the power balance across the transformer, and no voltage drop on the resonant tank, the RMS value of the high frequency resonant tank current is determined by battery charing power request:

\begin{equation}
I_{pri}^{RMS} = \frac{P_{batt}}{V_{pri}}
\end{equation}

According to the principle of operation, the only 4 transistors are conducting for any instance of time in the matrix converter (by pair in two commutated phase legs). Then the power dissipation in matrix converter is:

\begin{equation}
P_{loss}^{pri} = 4 \cdot (I_{pri}^{RMS})^2 R_{ds} = \frac{4 P_{batt}^2}{V_{pri}^2} R_{ds}
\end{equation}

For the secondary side which is essentially a synchronous rectifier, the power dissipation consists by 2 transistors (each per H-bridge leg) conducting RMS current corresponds to battery charging request: 

\begin{equation}
P_{loss}^{sec} = 2 I_{batt}^2 R_{ds} = \frac{2 P_{batt}^2}{V_{sec}^2} R_{ds}
\end{equation}

For $R_{ds} = 30$mOhm composite MOSFET device and requested output power $P_{batt} = 13$kW, the calculated losses are $P_{loss}^{pri} = 101.1$W and $P_{loss}^{sec} = 72.1$W, which with an assumed transformer losses $P_{loss}^{tx} = 90$W gives the efficiency of $98.0$\%.

\end{document}